\documentclass[conference]{IEEEtran}
\IEEEoverridecommandlockouts
\usepackage{cite}
\usepackage{amsmath,amssymb,amsfonts}
\usepackage{algorithmic}
\usepackage{graphicx}
\usepackage{booktabs}
\usepackage{hyperref}

\usepackage{textcomp}
\usepackage[dvipsnames]{xcolor}
\def\BibTeX{{\rm B\kern-.05em{\sc i\kern-.025em b}\kern-.08em
    T\kern-.1667em\lower.7ex\hbox{E}\kern-.125emX}}
\begin{document}
\bstctlcite{IEEEexample:BSTcontrol}

\title{MixFake: Benchmarking and Enhancing Audio Deepfake Detection in Diverse Real-world Mixed Audio\thanks{$^{\dagger}$Peng Cheng is the corresponding author (peng\_cheng@zju.edu.cn). This work was supported in part by the National Natural Science Foundation of China under Grant No. 62472372, and the Zhejiang Provincial Natural Science Foundation of China under Grant No. LD24F020010. }}

\author{
    \IEEEauthorblockN{
        Qingcao Li\textsuperscript{1,2}, 
        Yipeng Lin\textsuperscript{2}, 
        Weichen Lian\textsuperscript{2}, 
        Zhongjie Ba\textsuperscript{2,3}, 
        Peng Cheng\textsuperscript{2,3,$\dagger$}, 
        Zhichao Lian\textsuperscript{1}
    }
    \IEEEauthorblockA{
        \textsuperscript{1}School of Cyber Science and Engineering, Nanjing University of Science and Technology, Nanjing, China\\
        \textsuperscript{2}The State Key Laboratory of Blockchain and Data Security, Zhejiang University, Hangzhou, China\\
        \textsuperscript{3}Hangzhou High-Tech Zone (Binjiang) Institute of Blockchain and Data Security, Hangzhou, China\\
        Email: \{liqingcao, lzcts\}@njust.edu.cn, \{22521091, zhongjieba, peng\_cheng\}@zju.edu.cn, kevinbaylor@163.com
    }
}

\maketitle




\begin{abstract}
Speech deepfake detection has achieved remarkable success in clean environments but faces significant challenges in complex, real-world scenarios where speech is often mixed with background music or noise. Current state-of-the-art methods rely on semantic features from self-supervised learning (SSL) models, which often fail when processing non-speech or mixed-source audio. In this paper, we first introduce MixFake, a large-scale benchmark dataset designed to simulate diverse acoustic environments with varying SNR levels and mixed authenticity components. To address the "semantic-centric" limitation, we propose a Multi-stream Prompt Tuning framework that injects signal-level priors into SSL backbones. By integrating base, frequency, and texture streams through deep prompt injection, our model effectively captures acoustic artifacts. Experimental results demonstrate that our method significantly outperforms existing baselines, achieving a 0.95\% EER in foreground detection and a substantial 7.72\% absolute improvement in complex background detection tasks. Our dataset and code are available at \url{https://github.com/saltfish233/MixFake}.
\end{abstract}

\begin{IEEEkeywords}
Audio Deepfake Detection, Mixed Audio, Prompt Tuning, Hilbert-Huang transform, Teager-Kaiser energy operator
\end{IEEEkeywords}

\section{Introduction}
\label{sec:intro}

With the rapid development of speech synthesis techniques, Speech deepfake detection has emerged as a countermeasure resort given the potential harmful use of such methods. Speech synthesis, mainly text-to-speech (TTS) and voice conversion (VC), has been through decades of study. Text-to-speech synthesis is capable of generating an audio clip that sounds like a certain figure speaking arbitrary content, given a several-second audio recording of the target person and the corresponding text content~\cite{chen2025neural}. Modern neural VC systems can now transform a speaker's voice into a target identity with imperceptible latency, enabling seamless impersonation during live conversations~\cite{yang2024streamvc}. However, the easy access to these powerful tools increases the risks of voice impersonation, namely speech deepfake, and the corresponding consequences. To mitigate such risks, both industry and academia have been studying deep learning-based detection methods that differentiate bona-fide and synthesized speech signals~\cite{wang2020asvspoof, jung2022aasist, ba2023transferring}. 

Existing deepfake detection methods have made remarkable progress in terms of detection performance. These approaches prioritize improving the transferability of detection capability~\cite{tak2022rawboost}, which means the evaluation data is distributionally shifted away from the training data. This setting complies with practical scenarios, as it is impossible to predict the sources of forgery in speech samples. State-of-the-art (SOTA) techniques typically follow the paradigm of finetuning a Self-supervised Learning (SSL) foundation model with deepfake speech data, leveraging the vast speech knowledge learned by the pre-trained model to facilitate detection generalizability~\cite{tak2022automatic}. Continuous efforts focus on different feature extraction or more advanced classifiers to achieve continuous improvement in detecting out-of-training-domain deepfake samples, as evidenced by ASVspoof challenges~\cite{todisco2019asvspoof, liu2023asvspoof, wang2024asvspoof} and Speech Deepfake Arena Leaderboard~\cite{dowerah2025speech}.

Unlike previous studies \cite{yi2023audio}, we address the overlooked challenge of detecting deepfakes in complex audio. Standard benchmarks like ASVspoof \cite{wang2020asvspoof, jung2022aasist, ba2023transferring}, ADD \cite{yi2022add, yi2023add}, and In-the-wild \cite{muller2022does} predominantly feature clean speech with minimal ambient noise. In contrast, real-world scenarios involve diverse environments with multiple sound sources, such as background music. Crucially, existing methods often overlook manipulated background components—including synthesized ambient sounds or music—which creates a significant gap between laboratory conditions and practical deployment. Attackers can exploit this by embedding synthesized content into backgrounds or manipulating ambient noise to evade detection \cite{wu2024clad, muller2021speech, yi2024scenefake}.

Our evaluation of SOTA detection models reveals performance degradation on complex samples, even when utilizing fine-tuned pre-trained models. This is primarily because current self-supervised learning (SSL) frameworks \cite{baevski2020wav2vec, hsu2021hubert, chen2022wavlm} leverage pre-trained models for speech generation or translation tasks, which inherently prioritize semantic information \cite{pasad2021layer}. While effective for clean speech, this "semantic-centric" focus limits the model's ability to process background music or environmental sounds where linguistic context is absent.

To address this limitation, we propose a comprehensive benchmark dataset and a detection framework. Our dataset, MixFake, is built upon a systematic decoupling of speech, music, and environmental sounds, constructing a complete permutation matrix of genuine/fake combinations for both foreground and background components. Additionally, our Multi-stream Prompt Tuning framework incorporates signal-level acoustic priors into SSL backbones. The architecture integrates three collaborative streams: Base, Frequency, and Texture.

We specifically select the Hilbert-Huang Transform (HHT) \cite{huang1998empirical} to drive the Frequency stream because it is designed for analyzing non-stationary and non-linear audio signals by providing instantaneous frequency information\cite{jiang2019adaptive_src}. This helps our model capture subtle phase discontinuities and local frequency anomalies that characterize synthetic speech artifacts. For the Texture stream, we employ the Teager-Kaiser Energy Operator (TKEO) \cite{kaiser1990simple} due to its ability to track non-linear energy fluctuations with high temporal resolution\cite{LI2023119649}. This helps characterize ``texture'' differences between real signals and complex mixed-source forgeries across varying SNR levels. By injecting these multi-dimensional priors into every Transformer layer, our approach helps the model jointly leverage the foundation knowledge of the SSL backbone and signal-level acoustic artifacts.

Our contributions are summarized as following:
\begin{itemize}
    \item We propose MixFake, a benchmark dataset to facilitate audio deepfake research in mixed audio scenarios.
    \item We propose a multi-stream prompt tuning framework to inject signal-level priors into an SSL-based model, improving mixed-audio detection capability.
    \item Our comprehensive evaluations and ablation studies demonstrate the efficacy the proposed method and its key components.
\end{itemize}
\section{The MixFake Dataset}
\label{sec:mixfake_dataset}

\subsection{Overview}
\label{subsec:overview}


The \textbf{MixFake} dataset is a large-scale benchmark specifically designed to simulate the complexities of real-world acoustic environments for audio forgery detection. Unlike conventional datasets focused on single-source detection, MixFake encompasses both isolated sources and complex mixed-source scenarios. Specifically, these scenarios involve speech as the foreground and environmental sounds or music as the background, constructed through various permutations of bonafide and synthetic components. This design establishes a rigorous testbed for generalized synthetic speech detection.

As detailed in Table I, MixFake comprises 252,500 audio samples totaling approximately 673.69 hours, including 510.59 hours of single-source and 163.10 hours of mixed-source audios. To ensure robustness across diverse acoustic conditions, samples are uniformly distributed across SNR levels from -5dB to 20dB, preventing models from overfitting to specific energy ratios.We provide unified labels for single-source audio and granular, independent labels for foreground/background authenticity and SNR levels for mixtures. The dataset is partitioned into standard training, development, and evaluation sets.

\begin{table}[htbp]
\centering
\caption{Statistical distribution of the MixFake dataset.}
\label{tab:dataset_distribution}

\renewcommand{\arraystretch}{1.35} 

\setlength{\tabcolsep}{1.5pt} 

\resizebox{\columnwidth}{!}{
    \begin{tabular}{@{}l|cc|cc|cccc|c|c@{}}
    \toprule
    \textbf{Subset} & \multicolumn{2}{c|}{\textbf{Foreground Source}} & \multicolumn{2}{c|}{\textbf{Background Source}} & \multicolumn{4}{c|}{\textbf{Mixed}} & \textbf{Total} & \textbf{Dur.} \\ 
     & Real & Fake & Real & Fake & RF-RB & FF-RB & RF-FB & FF-FB & \textbf{Samples} & \textbf{(hrs)} \\ \midrule
    Train    & 2,000 & 10,000 & 4,000 & 20,000 & 20,000 & 20,000 & 20,000 & 20,000 & 116,000 & 371.02 \\
    Dev     & 500   & 1,000  & 1,000 & 2,000  & 4,000  & 4,000  & 4,000  & 4,000  & 20,500  & 43.67  \\
    Eval    & 2,000 & 10,000 & 4,000 & 20,000 & 20,000 & 20,000 & 20,000 & 20,000 & 116,000 & 258.99 \\ \midrule
    \textbf{Total} & \textbf{4,500} & \textbf{21,000} & \textbf{9,000} & \textbf{42,000} & \textbf{44,000} & \textbf{44,000} & \textbf{44,000} & \textbf{44,000} & \textbf{252,500} & \textbf{673.69} \\ \bottomrule
    \end{tabular}
}
\end{table}

Regarding synthesis diversity, the speech component integrates 19 distinct algorithms from \textbf{ASVspoof 2019 LA} \cite{wang2020asvspoof}. The music component features 10 non-overlapping algorithms derived from \textbf{Sonics} \cite{rahman2024sonics} and \textbf{FakeMusicCaps} \cite{comanducci2024fakemusiccaps}, while the environmental sounds leverage 3 representative algorithms from \textbf{EnvSDD} \cite{yin2025envsdd}.


\subsection{Dataset Construction Framework}
\label{subsec:framework}

\begin{figure*}[t]
    \centering
    \includegraphics[width=2.0\columnwidth]{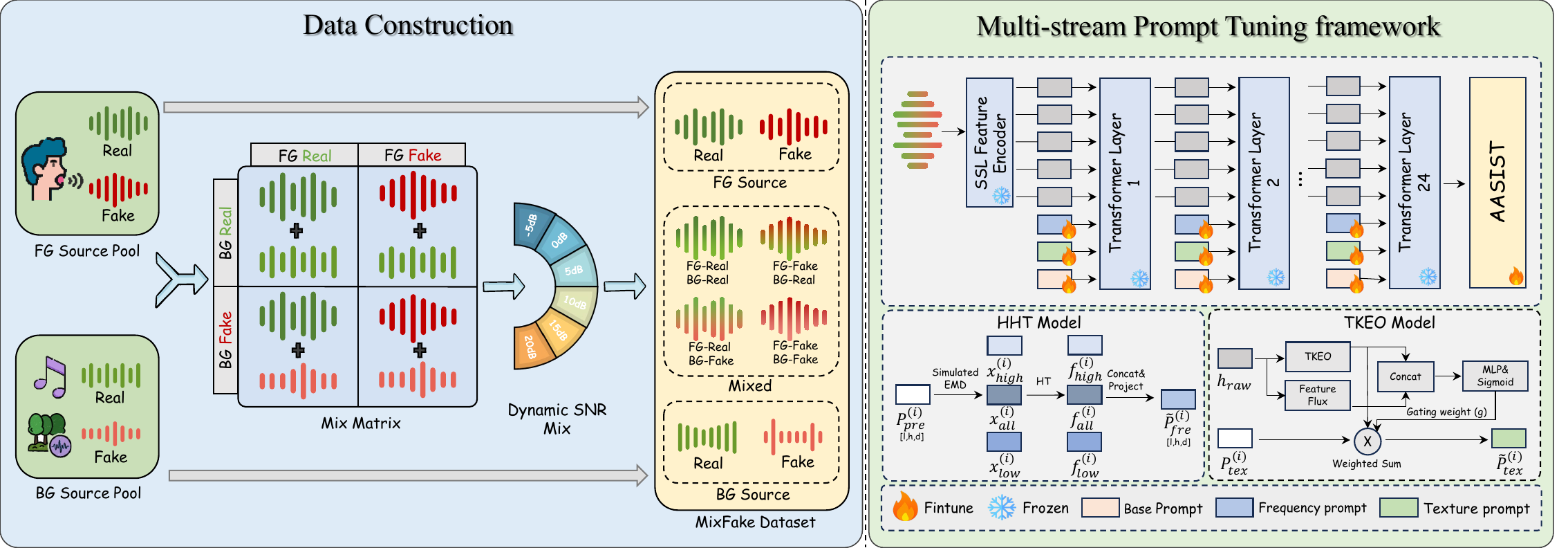}
    \caption{The overall framework of our proposed method. Left: The dataset construction pipeline for MixFake, highlighting the decoupled mixing strategy. Right: The Multi-stream Prompt Tuning framework, featuring the backbone model and deep prompt injection.}
    \label{fig:main_pic}
\end{figure*}

The construction of MixFake follows a modular pipeline consisting of two primary stages: source data selection and dynamic decoupled synthesis, as illustrated in the left panel of Fig. \ref{fig:main_pic}.


\subsubsection{Source Data Selection}
To maintain high category diversity, we established two independent source pools. The Foreground Source Pool is exclusively derived from ASVspoof 2019 LA, providing real and synthetic speech. The Background Source Pool utilizes a multi-source fusion strategy: \textit{Real backgrounds} aggregate music from FMA-Medium\cite{defferrard2016fma} and environmental sound (e.g., airport, street) from EnvSDD; \textit{Synthetic backgrounds} incorporate generative data from Sonics, FakeMusicCaps, and EnvSDD.

\subsubsection{Mixed Data Construction}
Once the source data is curated, we implement a dynamic synthesis process characterized by three key steps:
(1) \textbf{Authenticity Cross-Pairing:} Foreground and background sources are paired across all combinations (Real-Real, Fake-Real, Real-Fake, and Fake-Fake). With $mix\_ratio = 4$, each foreground sample is matched with four unique backgrounds to promote environment-agnostic learning.
(2) \textbf{Dynamic SNR Mixing:} Background gain is adjusted via RMS energy to match target SNRs randomly sampled from $\{-5, 0, 5, 10, 15, 20\}$~dB, maintaining energy consistency.
(3) \textbf{Temporal Alignment:} Background segments are looped or truncated to match the foreground duration, focusing the detection task on the primary signal.

\subsubsection{Single-Source Data Construction}
To provide benchmarks for isolated component detection and facilitate multi-scenario evaluation, MixFake incorporates single-source audio recordings. These samples are extracted directly from the source pools without mixing and include pure speech (foreground) as well as isolated music or environmental sounds (background). This subset allows the model to learn baseline acoustic characteristics for real and synthetic signals in isolation before confronting mixed-source complexities.

\section{Proposed Methodology}

The Multi-stream Prompt Tuning framework is illustrated in the right panel of Fig. 1. To enhance detection in diverse acoustic environments, we extend the XLSR-AASIST baseline by introducing signal-level analysis via deep prompt tuning. While pre-trained SSL models leverage high-level semantic features, these may be insufficient for capturing acoustic artifacts in mixed-source scenarios. We incorporate the Hilbert-Huang Transform (HHT) and Teager-Kaiser Energy Operator (TKEO) to provide a multi-dimensional signal-processing perspective for artifact identification.

\subsection{Backbone Model}

We employ XLSR-AASIST as the backbone network, the core of which is an SSL pre-trained Encoder: XLS-R. The architecture utilizes stacked Transformer blocks for layer-wise modeling of the temporal and frequency correlations within the audio sequence.

\subsection{Multi-stream Deep Prompt Injection Architecture}
The Multi-stream Deep Prompt Injection Architecture serves as the system's central axis, designed to address the specific complexities of the MixFake dataset. To mitigate the ``semantic-centric'' limitation of SSL models when processing mixed-source audio---where forged artifacts may reside in non-speech background components, we define three functionally distinct streams of learnable prompt embeddings at each layer $i$:
\begin{itemize}
    \item \textbf{Base Stream ($P_{base}^{(i)}$):} To provide a foundational learnable reference for the SSL backbone, these prompts are directly concatenated to the features without additional signal transformations. This stream maintains the baseline learnable capacity of the prompt tuning framework while allowing the other streams to provide specialized signal-level priors.
    \item \textbf{Frequency Stream ($\tilde{P}_{fre}^{(i)}$):} To capture fine-grained forgery traces often masked by background noise or music, this stream utilizes the HHT module to inject instantaneous frequency (IF) anomaly information, enabling the model to detect artifacts regardless of the source's semantic content.
    \item \textbf{Texture Stream ($\tilde{P}_{tex}^{(i)}$):} To distinguish between single-source and multi-source environments and adapt to varying SNR levels, this stream employs the TKEO module to modulate prompts using acoustic cues. By tracking energy distribution and feature flux from the input sequence, the module incorporates non-linear energy fluctuations into the learnable embeddings.
\end{itemize}

In each layer, the processed three-stream prompts are concatenated (denoted by the symbol $[ ; ]$) with the input audio features $H^{(i)}$ of that layer to form the actual input $X^{(i)}$ for the Transformer:
\begin{equation}
X^{(i)} = [P_{base}^{(i)} ; \tilde{P}_{fre}^{(i)} ; \tilde{P}_{tex}^{(i)} ; H^{(i)}]
\end{equation}
This layer-wise injection mechanism ensures that signal-level acoustic priors collaborate with the foundational prompts and pre-trained SSL representations, enabling the framework to generalize across the diverse authenticity combinations present in MixFake.

\subsection{Frequency Stream Module: Multi-scale HHT \& IF} \label{subsec:hht}
Corresponding to the Frequency Stream branch, we utilize a simulated multi-scale HHT module to process the frequency prompt vectors $P_{fre}^{(i)}$. The module first decomposes the prompt vector sequence into three components with distinct acoustic meanings $x_{scale}^{(i)}(n)$:
\begin{itemize}
    \item \textbf{High-frequency Component ($x_{high}^{(i)}$):} Extracted via temporal differencing to capture simulated high-frequency synthesis artifacts and phase jumps.
    \item \textbf{All-Frequency Component ($x_{all}^{(i)}$):} The prompt vector itself, retaining the original learnable features as a global reference.
    \item \textbf{Low-frequency Component ($x_{low}^{(i)}$):} Extracted via average pooling to capture the smooth overall trend of the signal.
\end{itemize}

For each component $x_{scale}^{(i)}(n)$ (where $scale \in \{high, all, low\}$), we extract the corresponding IF features through a unified process. First, an analytic signal $z^{(i)}(n)$ is constructed using the Hilbert Transform:
\begin{equation}
z^{(i)}(n) = x_{scale}^{(i)}(n) + j \hat{x}_{scale}^{(i)}(n)
\end{equation}
where $\hat{x}_{scale}^{(i)}(n)$ is the Hilbert transform of the signal. Based on the analytic signal, the instantaneous phase $\theta^{(i)}(n)$ at each time step $n$ is calculated:
\begin{equation}
\theta^{(i)}(n) = \arctan \left( \frac{\text{Im}[z^{(i)}(n)]}{\text{Re}[z^{(i)}(n)]} \right)
\end{equation}
where $\text{Re}[\cdot]$ and $\text{Im}[\cdot]$ denote the real and imaginary parts of a complex number, respectively. Subsequently, the instantaneous frequency $f_{scale}^{(i)}(n)$ of the component is obtained by calculating the phase change between adjacent time steps:
\begin{equation}
f_{scale}^{(i)}(n) = | \theta^{(i)}(n) - \theta^{(i)}(n-1) |
\end{equation}

Through this mapping, we obtain three distinct types of features: \textbf{Transient IF} ($f_{high}^{(i)}$), \textbf{Global IF} ($f_{all}^{(i)}$), and \textbf{Trend IF} ($f_{low}^{(i)}$). To integrate these multi-scale acoustic cues, the three IF sequences are concatenated along the feature dimension and subsequently mapped back to the prompt embedding space through a linear projection layer, yielding the final frequency prompt $\tilde{P}_{fre}^{(i)}$:
\begin{equation}
\tilde{P}_{fre}^{(i)} = \text{Linear}([f_{high}^{(i)} ; f_{all}^{(i)} ; f_{low}^{(i)}])
\end{equation}

\subsection{Texture Stream Module: TKEO \& Feature Flux} \label{subsec:tkeo}
Corresponding to the Texture Stream branch, this module is responsible for distinguishing between single-source audio and complex mixed audio. It modulates the learnable texture prompt vectors $P_{tex}^{(i)}$ using the pre-trained feature sequence $\mathbf{h}_{raw}$ extracted from the feature encoder by calculating the TKEO energy and Feature Flux. The TKEO for a sequence $\mathbf{h}_{raw}(n)$ is defined as:
\begin{equation}
\Psi[\mathbf{h}_{raw}(n)] = \mathbf{h}_{raw}^2(n) - \mathbf{h}_{raw}(n-1)\mathbf{h}_{raw}(n+1)
\end{equation}
where $\mathbf{h}_{raw}(n)$ denotes the feature vector of $H_{raw}$ at time step $n$. In our implementation, we calculate the absolute value $|\Psi[\mathbf{h}_{raw}(n)]|$ to track nonlinear energy fluctuations. Feature Flux is then calculated as the standard deviation of the absolute difference between adjacent vectors in the sequence:
\begin{equation}
\text{Flux} = \sigma(|\mathbf{h}_{raw}(n) - \mathbf{h}_{raw}(n-1)|)
\end{equation}

To handle the varying complexity across different samples, we implement an adaptive injection mechanism. A gating weight $g = \text{Sigmoid}(\text{MLP}([\bar{\Psi} ; \text{Flux}]))$ is computed, where $\bar{\Psi}$ denotes the mean TKEO energy across the sequence dimension. The final texture prompt $\tilde{P}_{tex}^{(i)}$ is obtained by fusing the learnable prompt embedding $P_{tex}^{(i)}$ with the acoustic energy cues:
\begin{equation}
\tilde{P}_{tex}^{(i)} = \text{LayerNorm}(g \cdot P_{tex}^{(i)} + (1 - g) \cdot \bar{\Psi})
\end{equation}

\subsection{Training Strategy}
To preserve the robust generalization capability of the pre-trained model, we freeze all original parameters of the SSL Encoder. The trainable parameters are limited to: 1) the three sets of multi-stream prompt vectors injected into each layer; 2) the signal analysis modules (HHT and TKEO); and 3) the backend classification network. This strategy enables the model to efficiently detect complex mixed audio with minimal parameter overhead.

\section{Experiments}
\label{sec:experiments}

\subsection{Experimental Settings}
\label{subsec:settings}

1) Datasets: Experiments primarily utilize the MixFake dataset for training and evaluation. For cross-dataset generalization, models are trained on the ASVspoof 2019 LA training set and evaluated on the In-the-Wild dataset.

2) {Baselines: To validate the advancement of our architecture, we selected three representative SOTA models:
(1) \textbf{XLSR-AASIST}~\cite{tak2022automatic}, a classic end-to-end framework combining XLS-R~\cite{babu2021xlsr} with a Graph Attention Network (GAT);
(2) \textbf{XLSR-Mamba}~\cite{xiao2025xlsr}, which integrates XLS-R with the long-sequence modeling of the Mamba SSM~\cite{gu2024mamba};
(3) \textbf{WPT-XLSR-AASIST}~\cite{xie2025detect}, a variant employing a prompt-tuning strategy.

3) Metrics: Equal Error Rate (EER) [33] is employed as the primary evaluation metric to assess classification performance. EER is chosen because it is the standard metric in the field of audio anti-spoofing and deepfake detection, providing a balanced measurement of the trade-off between False Acceptance Rate (FAR) and False Rejection Rate (FRR). Note that a lower EER value indicates better system performance.

4) Implementation Details: Audio samples are resampled to 16 kHz and 
to 4 seconds via padding or random cropping. Optimization is performed using the AdamW optimizer with a learning rate of $5 \times 10^{-3}$, a weight decay of $5 \times 10^{-4}$, and a batch size of 32. Training is conducted for 30 epochs on an NVIDIA H800 GPU.

\subsection{Detection Performance in Complex Scenarios}

\begin{table}[htbp]
    \centering
    \caption{Performance comparison on MixFake sub-tasks (EER \%).}
    \label{tab:main_exp_performance_comparison}
    \begin{tabular}{lcc}
        \toprule
        \textbf{Models} & \textbf{Foreground} & \textbf{Background} \\
        \midrule
        XLSR-AASIST     & 2.84\% (+1.89) & 20.12\% (+7.72) \\
        XLSR-Mamba      & 1.37\% (+0.42) & 17.86\% (+5.46) \\
        WPT-XLSR-AASIST & 2.85\% (+1.90) & 15.81\% (+3.41) \\ 
        \midrule
        \textbf{OURS}   & \textbf{0.95\%} & \textbf{12.40\%} \\
        \bottomrule
    \end{tabular}
\end{table}

To comprehensively evaluate our models' performance in complex acoustic environments, we designed two sub-tasks on the MixFake dataset:

\begin{itemize} 
\item \textbf{Foreground Speech Detection:} This task focuses on detecting the authenticity of foreground speech, treating background components as interference. Ground-truth labels are determined solely by the speech component's authenticity.
\item \textbf{Background Audio Detection:} Conversely, this task identifies the authenticity of background audio while treating foreground speech as interference. Ground-truth labels reflect the authenticity of the background component. 
\end{itemize}

The experimental results are summarized in Table~\ref{tab:main_exp_performance_comparison}. In the foreground speech detection task, our proposed method achieves SOTA performance, yielding an EER of 0.95\%. In contrast, representative baselines such as XLSR-AASIST and WPT-XLSR-AASIST exhibit significantly higher EERs of 2.84\% and 2.85\%, respectively. Even the competitive XLSR-Mamba (1.37\%) is outperformed by our method by a notable margin.

The background audio detection task proves to be far more challenging, as environmental sounds and music lack the explicit linguistic semantics that SSL-based models typically exploit. Consequently, all baseline models experience severe performance degradation; notably, the EER of XLSR-AASIST surges to 20.12\%. This confirms the inherent limitation of standard SSL backbones, which struggle to verify non-speech components due to their semantic-centric pre-training. Conversely, our method maintains a decisive advantage, achieving an EER of 12.40\%. This marks a substantial absolute improvement of 7.72\% and 5.46\% over XLSR-AASIST and XLSR-Mamba, respectively.

\subsection{Generalization Analysis}
\label{subsec:generalization}

To evaluate the generalizability of our proposed method in unseen scenarios, we conducted cross-dataset validation experiments. We trained the baseline models and our proposed method on the ASVspoof 2019 LA dataset and evaluated them on the In-the-wild dataset.

\begin{table}[htbp]
    \centering
    \caption{Experimental Results on In-the-wild Dataset (EER \%).}
    \label{tab:inthewild_results}
    \setlength{\tabcolsep}{25pt}      
    \begin{tabular}{lc}
        \toprule
        \textbf{Models} & \textbf{In-the-wild} \\
        \midrule
        XLSR-AASIST     & 9.60\% (+3.36) \\
        XLSR-Mamba      & 6.71\% (+0.47) \\
        WPT-XLSR-AASIST & 7.35\% (+1.11) \\
        \midrule
        \textbf{OURS}   & \textbf{6.24\%} \\
        \bottomrule
    \end{tabular}
\end{table}

The experimental results indicate that our proposed method exhibits superior generalization performance in cross-dataset validation. As illustrated in Table~\ref{tab:inthewild_results}, existing baseline models suffer from varying degrees of performance degradation when confronted with the out-of-domain In-the-wild test set. In contrast, our method achieved the lowest EER of 6.24\%. For comparison, the EERs for XLSR-AASIST, XLSR-Mamba, and WPT-XLSR-AASIST were 9.60\%, 6.71\%, and 7.35\%, respectively.

\subsection{Robustness Analysis}

To evaluate robustness across diverse acoustic environments, SNR tests were conducted on the mixed-source audio samples of the MixFake dataset ranging from $-5$~dB to $20$~dB. The results, illustrated in Fig. \ref{fig:robustness_results}, indicate that our method achieves a lower EER than SOTA baselines in both additive interference and signal masking scenarios, demonstrating superior stability.

\begin{figure}[htbp]
    \centering
    \includegraphics[width=0.9\columnwidth]{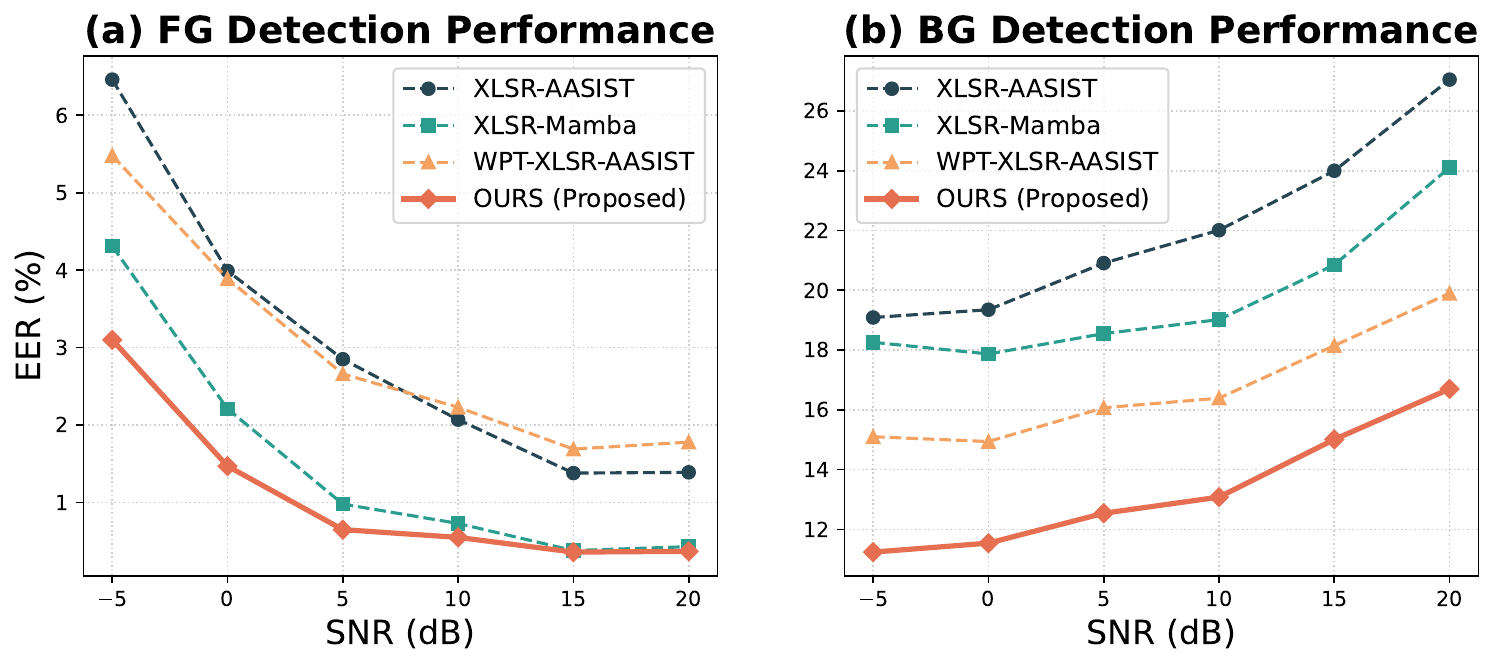}
    \caption{Performance comparison of baseline models and our proposed method under varying SNRs on the MixFake dataset.}
    \label{fig:robustness_results}
\end{figure}

In the foreground speech detection task (see Fig. \ref{fig:robustness_results}a), detection performance improves as the background interference diminishes (i.e., as SNR increases), with our method reaching an EER of $0.36\%$ at $15$~dB. At the $-5$~dB SNR level, where the loud background component poses the greatest interference, our method maintains an EER of $3.10\%$, outperforming the XLSR-AASIST ($6.46\%$) and WPT-XLSR-AASIST ($5.48\%$) baselines. Conversely, background audio detection (see Fig. \ref{fig:robustness_results}b) is most effective when the target background signal is more prominent, attaining its lowest EER of $11.24\%$ at $-5$~dB. As the background signal becomes relatively weaker and masked by dominant foreground speech at $20$~dB, the EER of the XLSR-AASIST baseline rises to $27.05\%$. In contrast, our method limits this value to $16.70\%$, indicating that the multi-stream architecture facilitates more effective signal disentanglement when background artifacts are obscured.

\subsection{Ablation Study}
\label{subsec:ablation}

\begin{table}[htbp]
\caption{Ablation results on the MixFake dataset (EER \%).}
\label{tab:ablation_results}
\begin{center}
\begin{tabular}{lcc}
\toprule
\textbf{Prompt Variants} & \textbf{Foreground} & \textbf{Background} \\
\midrule
$P_{base}$                  & 3.05\% (+2.10) & 14.31\% (+1.91) \\
$\tilde{P}_{fre}$          & 2.01\% (+1.06) & 13.50\% (+1.10) \\
$\tilde{P}_{tex}$          & 2.13\% (+1.18) & 14.89\% (+2.49) \\
$\tilde{P}_{tex} + P_{base}$ & 1.71\% (+0.76) & 13.62\% (+1.22) \\
$\tilde{P}_{fre} + P_{base}$ & 1.50\% (+0.55) & 12.86\% (+0.46) \\
$\tilde{P}_{tex} + \tilde{P}_{fre}$ & 1.35\% (+0.40) & 13.10\% (+0.70) \\
\midrule
\textbf{Ours ($P_{base} + \tilde{P}_{fre} + \tilde{P}_{tex}$)} & \textbf{0.95\%} & \textbf{12.40\%} \\
\bottomrule
\end{tabular}
\end{center}
\end{table}

Table~\ref{tab:ablation_results} summarizes the performance of various prompt configurations. Here, $P_{base}$, $\tilde{P}_{fre}$, and $\tilde{P}_{tex}$ represent variants where only the respective single stream is employed. Specifically, $P_{base}$ serves as the baseline, utilizing only the direct learnable prompts without any signal-level priors.

The results show that multi-stream configurations consistently outperform single-stream variants. In foreground detection, the individual signal-informed streams $\tilde{P}_{fre}$ ($2.01\%$) and $\tilde{P}_{tex}$ ($2.13\%$) both surpass the $P_{base}$ baseline ($3.05\%$). This confirms that HHT-based frequency and TKEO-based energy features provide critical information that complements the foundational prompts. For the more challenging background detection task, which lacks linguistic semantics, the HHT-based $\tilde{P}_{fre}$ alone reduces the EER to $13.50\%$. The optimal performance of $12.40\%$ is achieved through the full three-stream integration, indicating that signal-level priors are essential for identifying artifacts in non-speech components.
\section{Conclusion} \label{sec:conclusion} This paper investigates audio deepfake detection in complex mixed-source environments. We introduce MixFake, a large-scale benchmark designed to bridge the gap between laboratory-controlled speech and real-world audio complexity. To address the ``semantic-centric'' limitations of SSL models, we propose a Multi-stream Prompt Tuning framework integrating a Base Stream of learnable prompts with HHT-based frequency and TKEO-based texture priors. Experimental results demonstrate that this multi-dimensional approach achieves a 0.95\% EER in foreground detection and a 7.72\% absolute improvement in background detection, while maintaining robust cross-dataset generalization. Our work offers a new perspective for developing countermeasures effective in diverse, noisy acoustic scenes.

\bibliographystyle{IEEEtran}
\bibliography{icme2026references}

\end{document}